\documentclass{nature}
\usepackage[super]{natbib}
\usepackage{lineno,hyperref}
\usepackage[lined, boxed, linesnumbered, commentsnumbered]{algorithm2e}
\usepackage{algorithmic}
\usepackage{amsmath, amsthm, amssymb}
\usepackage{amsmath,bm,amsfonts}
\usepackage{graphicx}
\usepackage{subfigure}
\usepackage{color}
\usepackage{xcolor}

\makeatletter
\let\saved@includegraphics\includegraphics
\AtBeginDocument{\let\includegraphics\saved@includegraphics}
\renewenvironment*{figure}{\@float{figure}}{\end@float}
\makeatother

\title{Connecting shear localization with the long-range correlated	polarized stress fields in granular materials}

\author{Yinqiao Wang$^{1}$, Yujie Wang$^{1}$ \& Jie Zhang$^{1,2,3,*}$}

\begin{document}

\maketitle

\begin{affiliations}
	\item School of Physics and Astronomy, Shanghai Jiao Tong University, 800 Dong Chuan Road, Shanghai 200240, China
	\item Institute of Natural Sciences, Shanghai Jiao Tong University, Shanghai 200240, China
	\item Collaborative Innovation Center of Advanced Microstructures, Nanjing 210093, China
	\\ * Email: jiezhang2012@sjtu.edu.cn	
\end{affiliations}

\begin{abstract}
	One long-lasting puzzle in amorphous solids is shear localization, where local plastic deformation involves cooperative particle rearrangements in small regions of a few inter-particle distances, self-organizing into shear bands and eventually leading to the material failure. Understanding the connection between the structure and dynamics of amorphous solids is essential in physics, material sciences, geotechnical and civil engineering, and geophysics. Here we show a deep connection between shear localization and the intrinsic structures of internal stresses in an isotropically jammed granular material subject to shear. Specifically, we find strong (anti)correlations between the micro shear bands and two polarized stress fields along two directions of maximal shear. By exploring the tensorial characteristics and the rotational symmetry of force network, we reveal that such profound connection is a result of symmetry breaking by shear. Finally, we provide the solid experimental evidence of long-range correlated inherent shear stress in an isotropically jammed granular system.
\end{abstract}

\section*{Introduction}
Shear localization, in which local particle rearrangements appear in narrow regions of a few inter-particle distances, is a fascinating feature that not only appears in granular materials \cite{amon12hotSpots,lebouil14granular,denisov16granularAvalanch,zhengjie18} but also is shared by many other amorphous solids \cite{nicolas18review}, such as molecular glass\cite{albaret16}, metallic glass \cite{wang12metallicGlass}, colloids\cite{chikkadi11strainCorrelation,jensen14quadrupole,illing16supercooledLiquids}, emulsions \cite{desmond15emulsions} and foams \cite{kabla09Foam}. 
It is not only crucial to material research but also vital to the catastrophic failure of soils in geotechnical and civil engineering \cite{terzaghi96soilMechanics}, and the control of geo-hazards \cite{scholz19earthquakeBook}. Shear localization in amorphous solids remains under intense debate due to the disordered nature of materials; its ubiquity requires a general explanation. 
One important issue is regarding the initial shear localization when a homogeneous and isotropic amorphous solid is subject to shear. In particular, experiments on sand \cite{lebouil14granular} and two-dimensional (2D) granular materials \cite{zhengjie18} show that when a strain much smaller than the yield strain is applied, it appears immediately, showing self-organized spatial structures. The nature of the cooperation remains elusive.

The cooperative particle rearrangement implies long-range correlations, which are incorporated in microscopic theories \cite{langer01STZ,maloney06,tanguy06,dasgupta12energyMinimization} in two different means.
Both theories assume that local plastic particle rearrangements act like Eshelby inclusions \cite{eshelby57}, causing anisotropic and long range effects to their surrounding elastic media. The first theory is formulated on a dynamical basis: an avalanche of Eshelby inclusions leads to the shear-band formation \cite{langer01STZ,maloney06,tanguy06}. 
However, the recent granular experiment \cite{zhengjie18} shows no compelling evidence of the correlation between local stress drop and particle rearrangement, raising concerns of the relevance of the theory to granular materials, especially at the beginning of shear.    
The second theory is a mean-field theory based on an energy-minimization principle \cite{dasgupta12energyMinimization}. However, it is unclear how the force chains and dissipation in granular materials would corporate with the theory. 
On the other hand,  numerical works \cite{chowdhury16harrowell,lemaitre17inherentStress,maier17} observe that despite that the correlation of pressure is short range, the spatial correlation of inherent shear stress has an intrinsic quadrupolar anisotropy and a long-range power-law decay $\propto r^{-d}$ in $d$ dimensions, which is explained by a scaling argument assuming mechanical equilibrium and isotropy of amorphous solids \cite{lemaitre18theory}. Moreover, this observation is explained using field theories \cite{henkes2009fieldTheory,degiuli18PRE,degiuli18PRL}. However, experimental evidence is still missing. Besides, it is equally missing which role and to what extent the long-range stress correlation may play in the shear dynamics. Considering that the spatial correlation of stresses provides a quantification of force network in granular materials \cite{majmudar05nature,ostojic06forceNetwork}, it is natural to conjecture the missing role played by force network in the shear dynamics of isotropically jammed granular materials.

In this article, we address experimentally the connection between the shear localization and long-range correlated pseudo stress-chains at the beginning of shear. To this purpose, we study an isotropically jammed granular materials, consisting of bi-disperse photoelastic disks, subject to pure shear. We observe that micro shear bands occur right at the start of shear, generating those self-organized cooperative particle rearrangements. We find that there is a strong connection between the particle rearrangements and the pseudo stress-chains, as can be quantified by the correlation between the particle-rotation field and the polarized stress field $\tau_1(\alpha)$ along the directions of shear, $\alpha=\pm45^\circ$. 
Further analysis on $\tau_1(\alpha)$ for arbitrary angles $\alpha$ verify the infinite degeneracy of these long-range correlated polarized stress, showing a continuous rotational symmetry. Hence, we understand the deep connection between the dynamics of shear localization and the structure of pseudo stress-chains as symmetry breaking by shear.
Moreover, we observe that the chain-like characteristics of the polarized stress $\tau_1$ is quantitatively captured in the scaling anisotropy of its autocorrelation $\langle C_{\tau_1}(r,\theta)\rangle$, which shows a power law decay slower than $r^{-2}$ along the chain direction and faster than  $r^{-2}$ perpendicular to the chain direction. Surprisingly, the azimuthal averaged harmonic projection of the autocorrelation function 
\begin{equation}
\langle\bar{C}_{\tau_1}(r)\rangle = \pi^{-1}\int_0^{2\pi}\mathrm{d}\theta \cos(2\theta) \langle C_{\tau_1}(r,\theta)\rangle
\end{equation}
decays as a power law of $r^{-2}$, which is still consistent with the continuum description of the stress chains in the theories\cite{lemaitre18theory,henkes2009fieldTheory,degiuli18PRE,degiuli18PRL}.
Finally, we verify that the autocorrelations of inherent shear stress $\tau_2$ show a quadrupole-like pattern with a power law decay of $r^{-2}$, providing the experimental evidence of the theoretical predictions \cite{lemaitre18theory,henkes2009fieldTheory,degiuli18PRE,degiuli18PRL}.

\section*{Results}

\begin{figure}
	\centerline{\includegraphics[trim=0cm 0cm 0cm 0cm, width=0.8\linewidth]{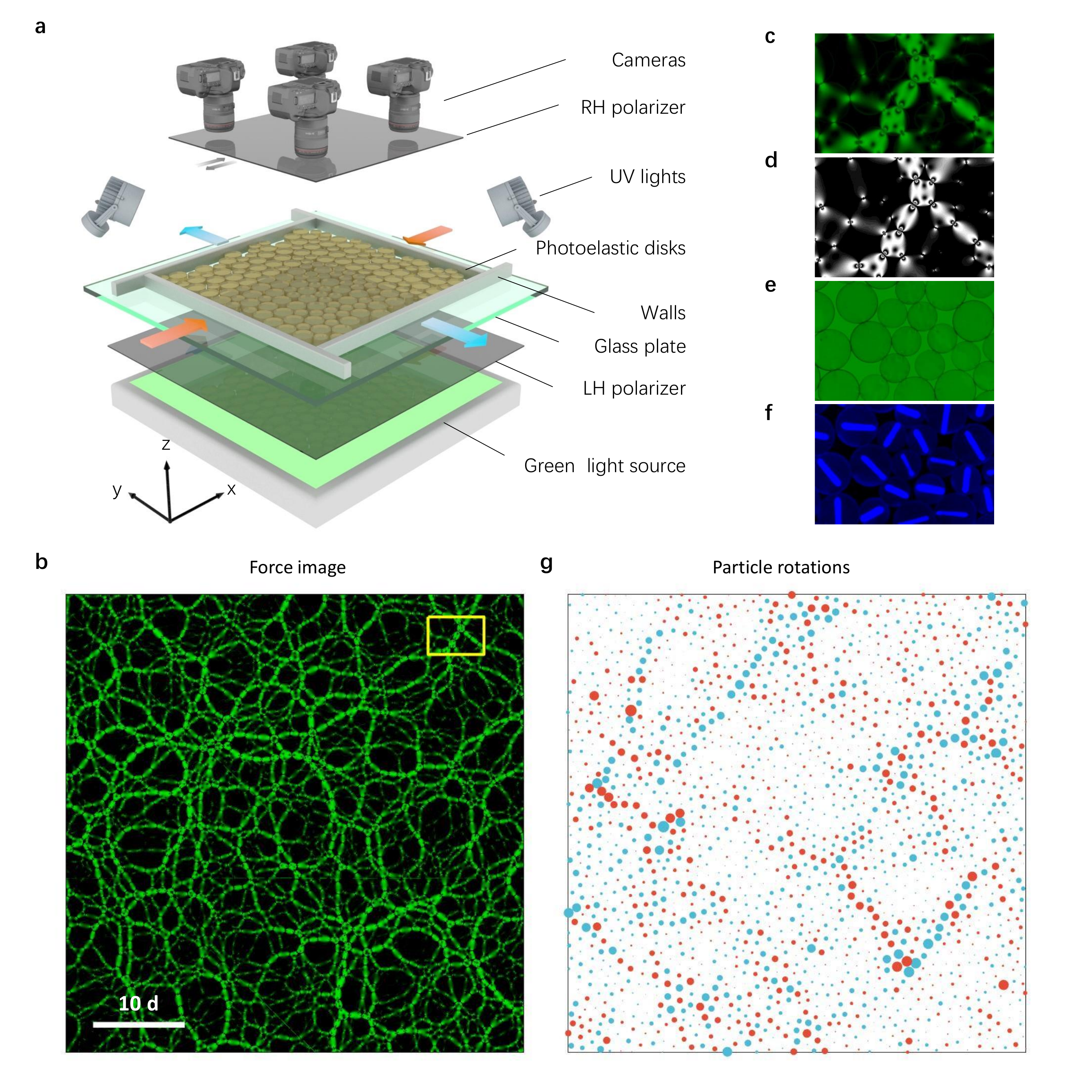}}
	\caption{\label{figure1}\textbf{Schematic of the experimental setup and different types of images recorded.} \textbf{a}, Experimental setup. \textbf{b}, A snapshot of force chains in an isotropically jammed state (an initial state before shear). A yellow rectangle draws a region that is amplified in panel \textbf{c}. \textbf{d}, a corresponding computer reconstructed image using the measured vector contact forces. \textbf{e}, a corresponding normal image of particle configuration. \textbf{f}, a corresponding UV image for tracking particle rotations. \textbf{g}, Spatial distribution of particle rotations, measured from the initial sate $\gamma = 0$ in panel \textbf{b} to $\gamma=0.75\%$. The sizes of red/blue circles are proportional to the magnitudes of counterclockwise/clockwise rotations. Scale bar = 10 $\mathrm{d}$, where $\mathrm{d=1\ cm}$, is the diameter of a small particle.}
\end{figure}

\subsection{Initial shear localization.}
We first prepare an isotropic jammed packing using a biaxial apparatus shown in Fig.~\ref{figure1}a, then apply pure shear quasi-statically in steps by compressing along $x$ axis and expanding along $y$ axis while keeping the area fixed. More experimental details can be seen in Methods. Surprisingly, shear localization appear at a strain $\gamma = 0.75\%$ much smaller than the yield strain $\gamma_{\rm{y}}\sim3\%$, as shown in Fig.~\ref{figure1}g. Here we use particle rotations to characterize shear localization, whose pattern is similar with local shear strain and $D^2_{\min}$ \cite{falk98langer}, as shown in the Supplementary Figure 1. 
Local shear strain describes the affine deformation of the nearest neighbors of a particle, and $D^2_{\min}$ quantifies the degree of the nonaffine displacement of the particle relative to its nearest neighbors by subtracting its affine displacement. 
Particles rotating clockwise tend to align cooperatively in bands along the $45^\circ$ direction, whereas those rotating counterclockwise tend to align in bands along the $-45^\circ$ direction, which leads to a quadrupole-like long-range correlator of local strain, as shown in the Supplementary Figure 2, consistent with previous experiments of colloids \cite{chikkadi11strainCorrelation,jensen14quadrupole,illing16supercooledLiquids} and 3D granular materials \cite{lebouil14granular}. 

In contrast, spatial distributions of local stress changes are rather homogeneous without obvious correlation with particle rearrangements, as shown in the Supplementary Figure 1.
As discussed in detail in the introduction, microscopic theories \cite{langer01STZ,maloney06,tanguy06, dasgupta12energyMinimization} can not fully explain the emergent behaviors of shear localization seen in Fig.~\ref{figure1}g, which shows organized spatial structures of particle rotation with long range correlation. It is a great challenge to understand the emergence of those self-organized, cooperative particle rearrangements in the seemingly disordered granular material. Recall that recent simulations and theories show the exhibition of long-range correlated shear stress of inherent structures in amorphous solids, including granular materials \cite{lemaitre18theory,henkes2009fieldTheory,degiuli18PRE,degiuli18PRL}, though no experimental evidence has been found yet, in particular, not in granular materials. If it does exist, such as in the force-chain network shown in Fig.\ref{figure1}b, it is urgent to unravel the possible connection between the force network structure and the particle dynamics.
 
 \subsection{Polarized stress fields.}
To reveal this connection, we first define the Cauchy stress tensor of individual particle in the initial state,
\begin{equation}
\begin{aligned}
\boldsymbol{\sigma}_i
\equiv \begin{bmatrix}
\sigma_{i,xx}&\sigma_{i,xy} \\ 
\sigma_{i,yx}&\sigma_{i,yy} 
\end{bmatrix} 
\equiv 
\frac{1}{S_i}\sum_j\boldsymbol{r}_{ij}\otimes\boldsymbol{f}_{ij} 
\end{aligned}
\end{equation}
Here $\boldsymbol{r}_{ij}$ is the position vector from the center of disk $i$ to the contact point between disks $i$ and $j$, $\boldsymbol{f}_{ij}$ is the contact-force vector between disks $i$ and $j$, $S_i$ is the area of the Voronoi cell of disk $i$, the operator '$\otimes$' represents the dyadic product of two vectors, and the index $j$ runs over all disks $j$ in contact with disk $i$.
In an arbitrary Cartesian axes $(x^\prime,y^\prime)$, of an angle $\alpha$ with respect to the laboratory axes $(x,y)$, the stress tensor is given by,
\begin{equation}
\begin{aligned}
\boldsymbol{\sigma}_{i}^{\prime}
&=\left[\begin{array}{cc}{\cos \alpha} & {\sin \alpha} \\ {-\sin \alpha} & {\cos \alpha}\end{array}\right] 
\boldsymbol{\sigma}_i
\left[\begin{array}{cc}{\cos \alpha} & {-\sin \alpha} \\ {\sin \alpha} & {\cos \alpha}\end{array}\right]
\end{aligned}
\end{equation}

\begin{figure}
	\centerline{\includegraphics[trim=0cm 0cm 0cm 0cm, width=1\linewidth]{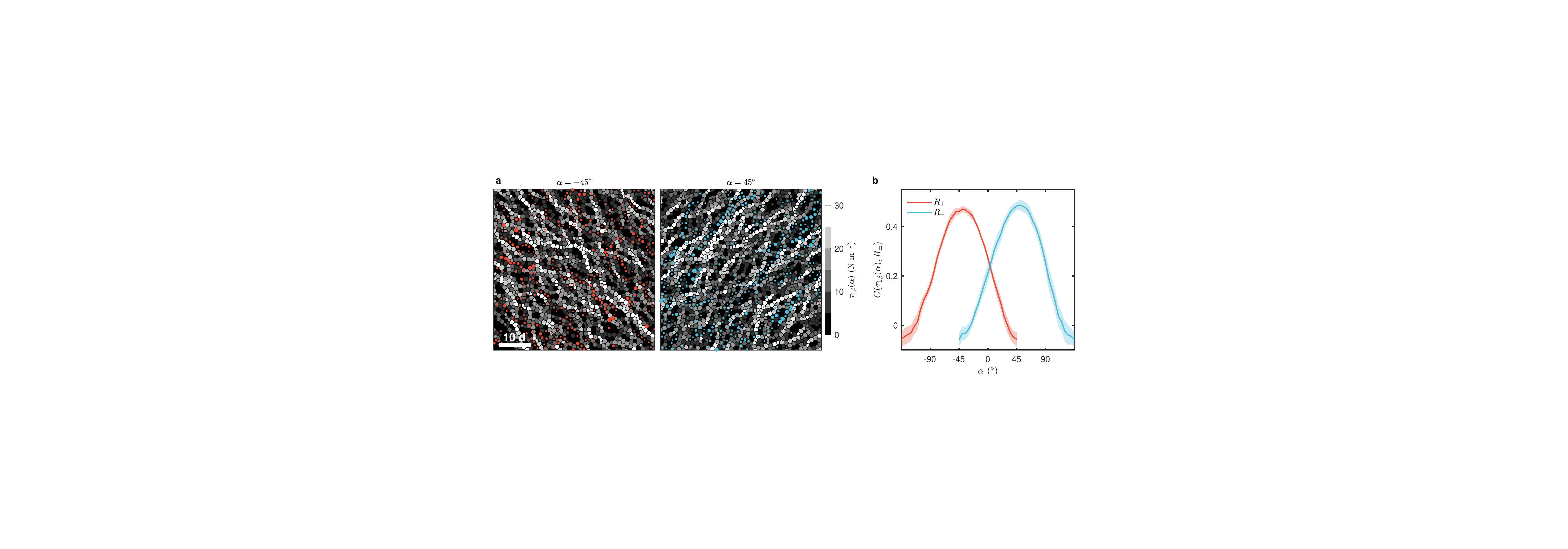}}
	\caption{\label{figure2}\textbf{Anti-correlation between particle rotations and polarized stress fields.} \textbf{a}, A superposition of the polarized stress fields $\tau_1(\alpha)$ with $\alpha=\pm 45^\circ$ and the counterclockwise(red circles)/clockwise(blue circles) particle rotations, the same as in Fig.~\ref{figure1}\textbf{g}. In background, the Voronoi cells of disks are painted according to the gray scales of $\tau_{1,i}(\alpha)$ of each disk $i$. Scale bar = 10 $\mathrm{d}$. \textbf{b}, The correlations $C(\tau_{1,i}(\alpha),R_{\pm})$ between particle rotations ($R_+$: counterclockwise, $R_-$: clockwise) and $\tau_{1,i}(\alpha)$ versus angle $\alpha$. The bands indicate standard errors of six independent runs.}
\end{figure}

\subsection{Correlation between structure and dynamics.}
Remarkably, we found strong correlations between the micro-bands and the spatial distributions of normal stress along $\pm 45^\circ$, as shown in Fig.~\ref{figure2}a. We denote the normal stress along $\alpha$ direction as the polarized stress $\tau_{1,i}(\alpha)=\sigma_{i, x^{\prime} x^{\prime}}$, showing chain-like structures, which we call pseudo stress-chains. We denote the field of $\tau_{1,i}(\alpha)$ of all disks $i$ as $\tau_1(\alpha)$.
The micro bands, i.e. those cooperative particle motions, just locate within the inter-spaces of pseudo stress-chains in the two polarized stress fields $\tau_1(\alpha)$ with angles $\alpha=\pm 45^\circ$ along directions of maximal shear.
At the same time, other quantities, including free volume, contact number and deviatoric shear stress, show weak correlations with the micro-bands, as shown in Supplementary Figure 3.
To quantify the correlations, we compute the correlation function $C(\tau_{1,i}(\alpha),R_{\pm})$ following Ref. \cite{patinet16localYieldStress}. We choose the particles of top $10\%$ (counterclockwise $R_+$ and clockwise $R_-$) rotations, then the median value $\tau_1^{{\rm{m}},R_{\pm}}(\alpha)$ of the polarized stress of these particles and the cumulative distribution functions (CDF) of $\tau_{1,i}(\alpha)$ give the correlation as $C(\tau_{1,i}(\alpha),R_{\pm}) = 1-2\cdot \textrm{CDF}(\tau_1^{{\rm{m}},R_{\pm}}(\alpha))$. The results are shown in Fig.~\ref{figure2}b, where the peak values are around $\alpha=\pm45^o$, as expected. In the above calculation of the correlation functions, changing the cutoff (top 10\%) of the particle rotation has little influence on the results, as shown in Supplementary Figure 4.

\begin{figure}
	\centerline{\includegraphics[trim=0cm 0cm 0cm 0cm, width=0.7\linewidth]{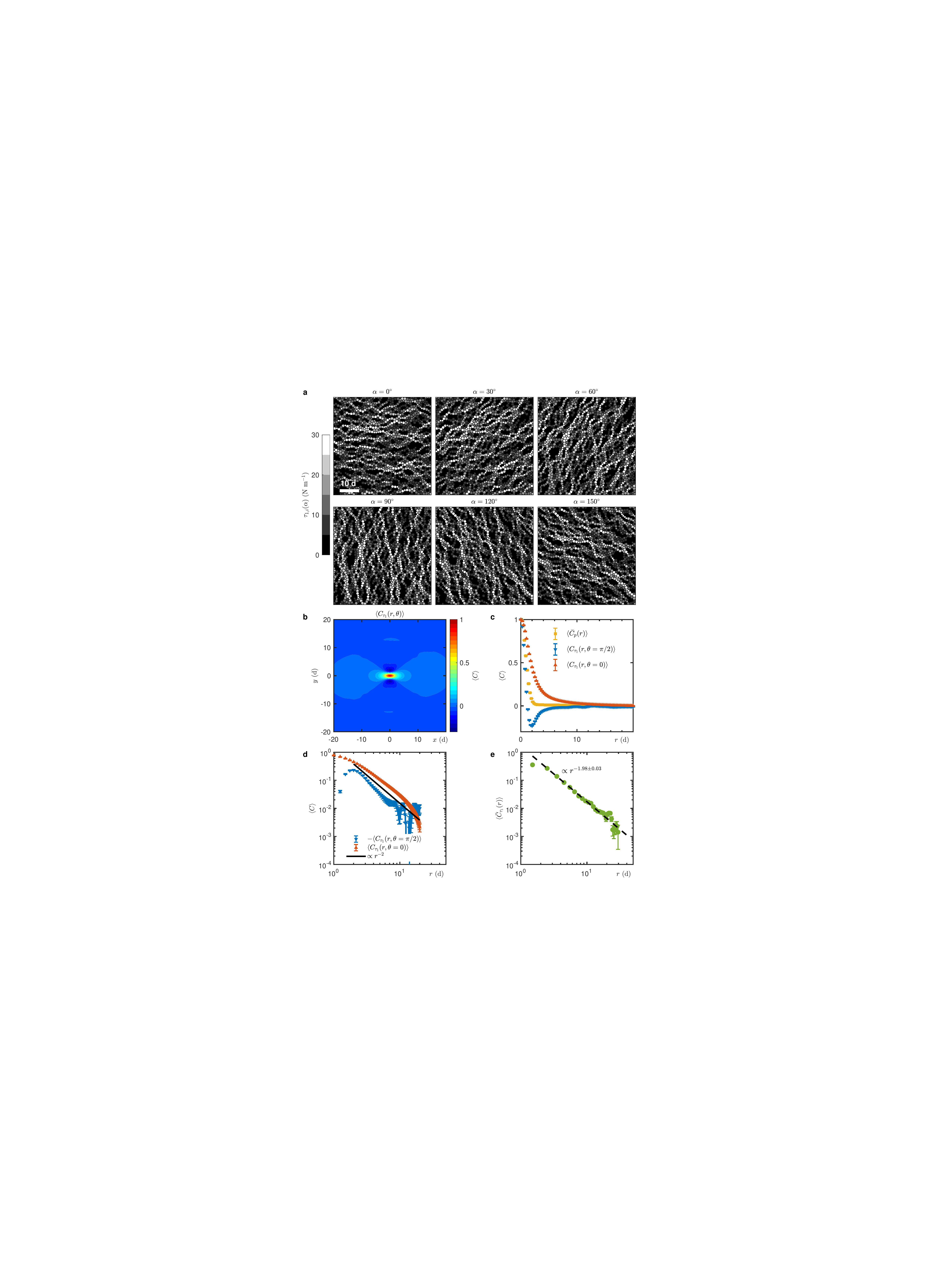}}
	\caption{\label{figure3} \textbf{Spatial distributions of polarized stress fields and their autocorrelation functions.} \textbf{a}, Spatial distributions of six polarized stress fields $\tau_1(\alpha)$. Scale bar = 10 $\mathrm{d}$. \textbf{b}, Spatial autocorrelation map $\langle C_{\tau_1}(r,\theta)\rangle$ of coarse grained polarized stress $\tau_{1}(\alpha)$, $\langle\cdots\rangle$ denotes ensemble average over 100 configurations and different polarized angles $\alpha$. Here, $\langle C\rangle$ denotes correlation functions. \textbf{c}, Cuts of $\langle C_{\tau_1}(r,\theta)\rangle$ along $\theta=0,\pi/2$ and autocorrelation function of coarse grained pressure $\langle \bar{C}_{p}(r)\rangle$, the over bar denotes average over $\theta$. 
	\textbf{d}, Log-log plots of $\langle C_{\tau_1}(r,\theta=0)\rangle$ and $-\langle C_{\tau_1}(r,\theta=\pi/2)\rangle$. The solid line $\propto r^{-2}$ is guide to the eye.
	\textbf{e}, The azimuthal averaged autocorrelation function of $\tau_1$, $\langle\bar{C}_{\tau_1}(r)\rangle$. The error bars represent the standard errors. The black dashed line indicates a power law fit of $\langle\bar{C}_{\tau_1}(r)\rangle$ for $r>3\ \mathrm{d}$.}
\end{figure}

\subsection{Symmetry and long-range characteristics of stress fields.}
From the symmetry perspective, the two polarized stress fields of $\tau_1(\alpha)$ along $\alpha=\pm 45^\circ$ are by no means special owing to the rotational symmetry of the isotropically jammed packing as shown in Fig.~\ref{figure1}b.
Six more polarized stresses  $\tau_1(\alpha)$ of $\alpha=0^\circ,30^\circ,60^\circ,90^\circ,120^\circ,150^\circ$ are shown in Fig.~\ref{figure3}a, showing filamentary, pseudo stress-chains preferentially aligned along the orientation of polarized angle $\alpha$. This long-range character is quantified using an autocorrelation function $C_{\tau_1}(r,\theta)$ of coarse-grained polarized stress \cite{lemaitre17inherentStress,goldhirsch02CG}, showing a dipolar signature in Fig.~\ref{figure3}b, similar with the pressure or force magnitude autocorrelations in shear-jammed granular systems \cite{majmudar05nature,lois2009}. Compared to the short-range correlation of local pressure, as shown in Fig.~\ref{figure3}c, $C_{\tau_1}(r,\theta)$ decays much slower along a cut in the dipolar direction. One prominent feature of the correlator of polarized stress is the anisotropy of the scaling property, as shown in the log-log plot in Fig.~\ref{figure3}d: the correlation function decays slower than $r^{-2}$ along the direction of the stress chains, i.e. $\theta = 0$; it decays much faster than $r^{-2}$ along the direction perpendicular to the direction of stress chains, i.e. $\theta= \pi/2$.
Nonetheless, a power-law $r^{-n}$ fit of $\langle\bar{C}_{\tau_1}(r)\rangle$, yields an exponent $n=1.98\pm 0.03$ as shown in Fig.~\ref{figure3}e, which indicates that an azimuthally averaged harmonic projection of the correlation function does show the scaling consistent with the field theoretical predictions of $n=2$ in 2D systems for the description of stress chains in the continuum limit \cite{lemaitre18theory,henkes2009fieldTheory,degiuli18PRE,degiuli18PRL}. A finite-size analysis of $\langle\bar{C}_{\tau_1}(r)\rangle$ is given in the Supplementary Figure 5a. 
By definition, $\langle \tau_{1,i}(\alpha) \rangle=(\sum_i \tau_{1,i}(\alpha)S_i)\cdot(\sum_i S_i)^{-1}=p$ for an isotropically jammed packing, where $S_i$ is the Voronoi area of disks $i$. Here $p$ is the pressure of whole system. 
Owing to the continuous rotational symmetry, there is an infinite degeneracy of such long-range correlated polarized stress fields $\tau_1(\alpha)$. Thus, the emergence of shear localization, i.e. micro bands along $\pm45^\circ$, can be understood as the breaking of the continuous rotational symmetry by shear.
 
\begin{figure}
	\centerline{\includegraphics[trim=0cm 0cm 0cm 0cm, width=0.8\linewidth]{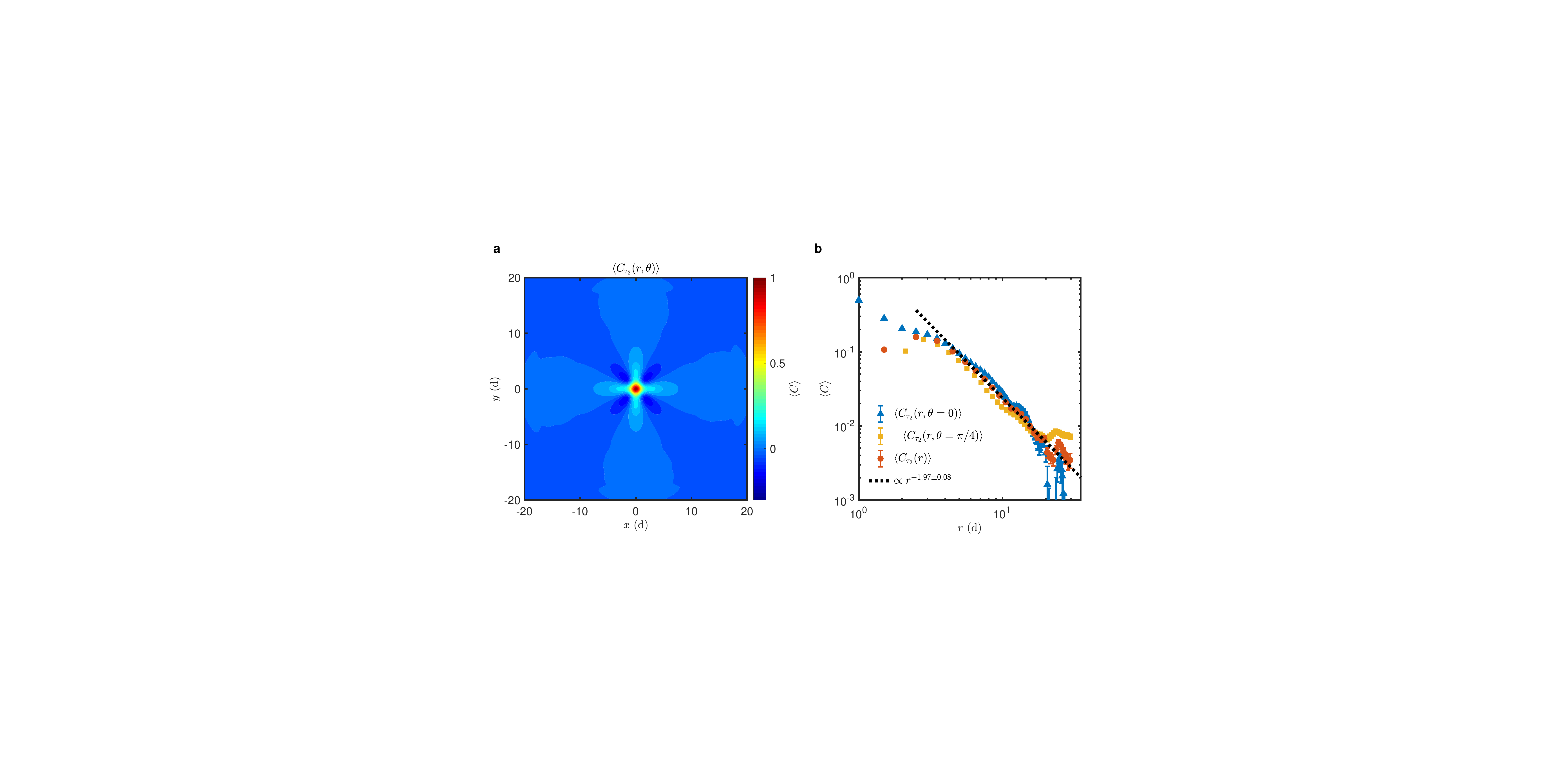}}
	\caption{\label{figure4}\textbf{Autocorrelation functions of shear stress $\tau_2$.} \textbf{a}, Spatial autocorrelation map $\langle C_{\tau_2}(r,\theta)\rangle$ of coarse grained shear stress $\tau_{2}(\alpha)$, $\langle\cdots\rangle$ denotes ensemble average over 100 configurations and different polarized angles $\alpha$. \textbf{b}, Cuts of $\langle C_{\tau_2}(r,\theta)\rangle$ along $\theta=0$ and $\theta=\pi/4$, and their azimuthal averaged autocorrelation function $\langle\bar{C}_{\tau_2}(r)\rangle$. The error bars represent the standard errors. The black dashed line indicates a power law fit of $\langle\bar{C}_{\tau_2}(r)\rangle$ for $r>6\ \mathrm{d}$.}
\end{figure}

Compared to the polarized stress $\tau_{1,i}(\alpha)$, in literature \cite{lemaitre18theory,henkes2009fieldTheory,degiuli18PRE,degiuli18PRL}, much attention has been paid to the quantity $\tau_{2,i}(\alpha)$, the so called inherent shear stress, due to the anisotropy and long-range characteristics in its spatial autocorrelation despite that no experimental evidence has been reported thus far \cite{lemaitre18theory,henkes2009fieldTheory,degiuli18PRE,degiuli18PRL}. First note that the $\tau_{2,i}(\alpha)$ is directly related to the polarized stress $\tau_{1,i}(\alpha)$ at individual particles by $\tau_{2,i}(\alpha)=\tau_{1,i}(\alpha)-\tau_{1,i}(\alpha+(\pi/2))$. Next, we present the spatial autocorrelation map of shear stress $\tau_{2,i}(\alpha)$, which indeed supports the theories \cite{lemaitre18theory,henkes2009fieldTheory,degiuli18PRE,degiuli18PRL}. The autocorrelation map clearly shows a quadrupole-like pattern with $\cos(4\theta)$ symmetry, as shown in Fig.~\ref{figure4}a, consistent with theories \cite{lemaitre18theory,henkes2009fieldTheory,degiuli18PRE,degiuli18PRL}. To verify the power-law decay of the correlation function, cuts along the $\theta=0,\pi/4$ lobes and an appropriate spherical harmonics projection
\begin{equation}
\langle\bar{C}_{\tau_2}(r)\rangle = \pi^{-1}\int_0^{2\pi}\mathrm{d}\theta \cos(4\theta) \langle C_{\tau_2}(r,\theta)\rangle
\end{equation} 
are shown in Fig.~\ref{figure4}b. A power-law $r^{-n}$ fit of $\langle\bar{C}_{\tau_2}(r)\rangle$ give the exponent $n=1.97\pm 0.08$, consistent with the theoretical prediction $n=2$ in 2D system \cite{lemaitre18theory,henkes2009fieldTheory,degiuli18PRE,degiuli18PRL}. A finite-size analysis of $\langle\bar{C}_{\tau_2}(r)\rangle$ is given in the Supplementary Figure 5b.
The deviation within $r\approx 6\ \mathrm{d}$, corresponds to the breakdown of continuum medium \cite{schirmacher13bosonPeak,wang18bosonPeak}.

\section*{Discussion}
To conclude, we find the collective particle rearrangements of the emergent behavior of shear localization in an isotropically jammed granular material are closely related to the pseudo stress-chains in the polarized stress fields $\tau_1(\alpha)$ for $\alpha$ along the directions of shear. The emergence of shear localization is associated with the breaking of the continuous rotational symmetry by shear. This mechanism is based on symmetry and the long-range character of internal stress without invoking the Eshelby mechanisms, which could also be applied to other amorphous solids. Statistically, the long-range characteristic of the internal stress are revealed from either the spatial correlations of polarized stress $\tau_1$ or the inherent shear stress $\tau_2$, which provides the direct experimental evidence of theoretical predictions \cite{lemaitre18theory,henkes2009fieldTheory,degiuli18PRE,degiuli18PRL}.
The present work serves as a starting point to understand more complicated dynamical processes of the evolution of shear localization.  
As strain increases, especially near the yielding, we suspect that Eshelby processes may contribute to the evolutions of shear localization and the development of global shear bands, which will be an important subject in future studies.   

\begin{methods}
	\subsection{Experimental details.}
In this experiment, we use a biaxial apparatus to apply isotropic compression or pure
shear on a two-dimensional granular system. The apparatus mainly consists of a rectangular frame mounted on top of a powder-lubricated glass plate with four walls that can move symmetrically with a motion precision of 0.1 mm while keeping the center of mass fixed. The basal friction coefficient is around 0.3. We estimate that the force magnitude of the basal friction is about $36$ times smaller compared to the typical contact-force magnitude. Hence the basal friction is negligible.
The rectangular area is filled with a random mixture of 2680  bi-disperse photoelastic disks (Vishay PSM-4) of with diameters of 1.4 cm and 1.0 cm and a number ratio of 1:1 to create various unjammed random initial configurations. The four narrow bands between these photoelastic disks and mobile walls are padded using a set of 300 small and 300 large bi-disperse Teflon-taped metal disks of the same sizes of the photoelastic disks. The friction coefficient is less than 0.1 between the Teflon-taped metal disks and the Teflon-taped mobile walls, which eliminates substantially the collective rotational motion of individual disks near the boundaries. Next, we apply isotropic compression to achieve packing at particular pressure levels.
To minimize the potential inhomogeneity of force chains in the jammed packing, we constantly apply mechanical vibrations in random directions of the horizontal plane before the packing fraction $\phi$ (the ratio between the area of disks and that of the system) exceeds the jamming point $\phi_J\approx 84.0\%$ of frictionless particles \cite{ohern03Jamming}. 
At the top, an array of 2 times 2 high-resolution (100 pixel per cm) cameras are aligned and synchronized. Figure~\ref{figure1}b shows one merged image of force-chain network of an isotropic jammed packing based on the pre-calibration of four cameras.
The packing in Fig.~\ref{figure1}b is confined in a square domain of 67.2 cm times 67.2 cm. Here,   
$\phi\approx 84.4\%$, the mean coordination number is around 4.1, and the pressure is around 11 N m$^{-1}$. 

We then apply pure shear quasi-statically in steps by compressing along $x$ axis and expanding along $y$ axis while keeping the area fixed. The step size is 0.5 mm, resulting the $\sim 0.15\%$ change of the strain.
At each step, three different images are recorded as shown in Fig.~\ref{figure1}. Disk positions are obtained using the normal image. Hough transformation is used to detect the particle position with a sub-pixel resolution. A UV image is taken for tracking individual particle rotation during shear, whose uncertainty is less than 0.02 rad. Contact forces are analyzed from the force-chain image using force-inverse algorithm, which generates a computed force-chain image based on an initial guess of contact forces, and then iterate contact forces to  minimize the difference between experimental and computed force-chain image \cite{majmudar05nature,daniels17photoelasticity,zadeh19}. The relative error of contact force measurement is around 3-4\% for the typical force magnitude, and the accuracy of contact forces is checked by plotting a computed image for comparison as shown in Fig.~\ref{figure1}d. 	

\subsection{Data availability.} The datasets generated during and/or analysed during the current study are available from the corresponding author on reasonable request.
\end{methods}

\begin{addendum}
	\item J.Z. acknowledges the support from National Natural Science Foundation of China under (No.11774221 and No. 11974238)
	\item[Author Contributions] Y.Q.W performed the experiment and carried out the data analyses. J.Z. designed the experimental project. Y.Q.W., Y.J.W. and J.Z. contributed in writing the manuscript and interpreting the data.
	\item[Competing Interests] The authors declare no competing interests.
	\item[Additional Information] Correspondence and requests for materials should be addressed to J.Z. (email: jiezhang2012@sjtu.edu.cn).
\end{addendum}

\clearpage
\setcounter{figure}{0}
\renewcommand{\thefigure}{S\arabic{figure}}
\section*{\Large{Supplementary Information}}

\begin{figure}[h!]
	\centerline{\includegraphics[trim=0cm 0cm 0cm 0cm, width=0.8\linewidth]{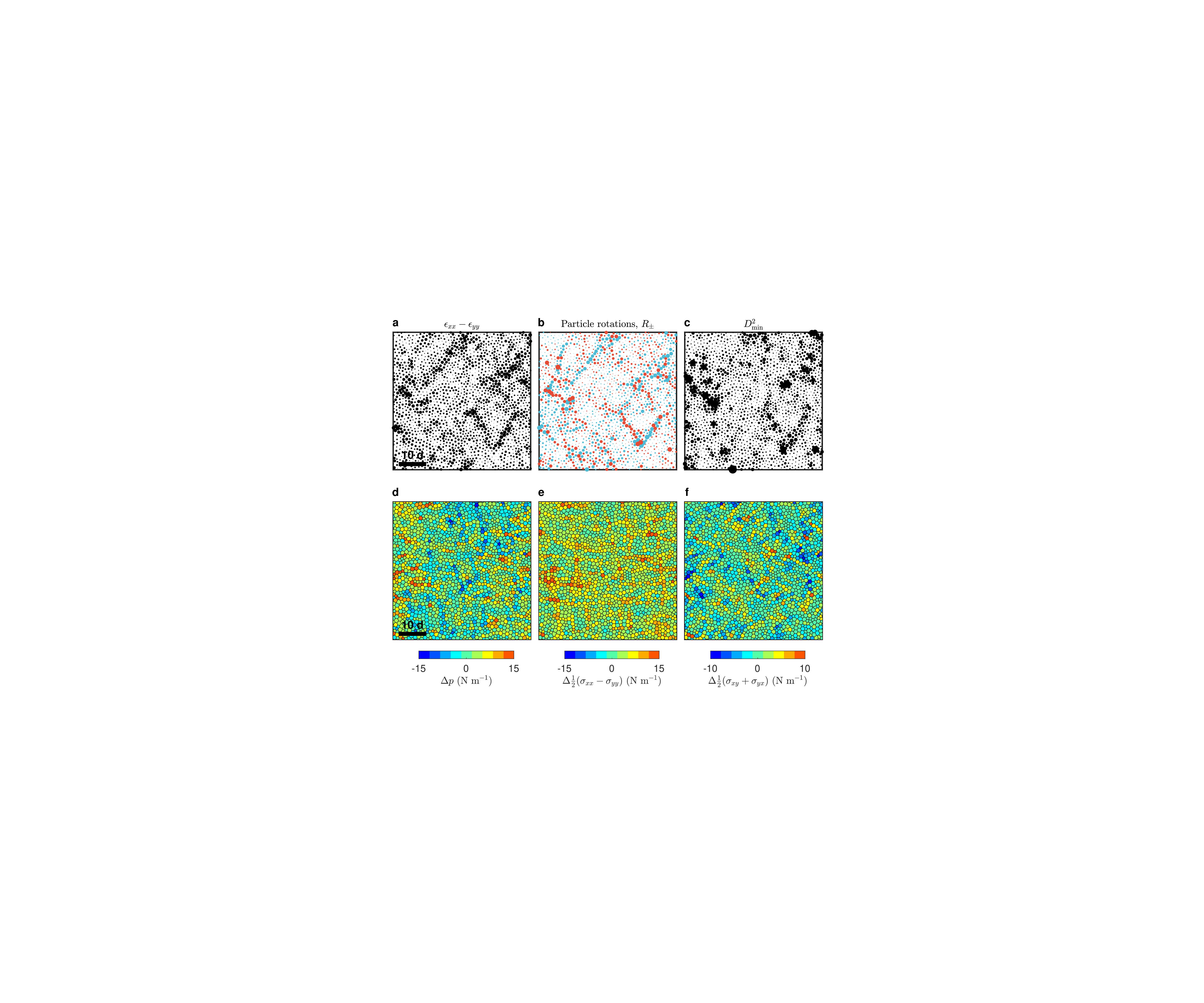}}
	\caption{\textbf{Shear localization and particle-scale stress changes.} Spatial distributions of \textbf{a}, local shear strain $\epsilon_{xx}-\epsilon_{yy}$, \textbf{b}, particles rotations (blue circles indicate clockwise rotations, red circles indicate counterclockwise rotations) and \textbf{c}, $D_{\mathrm{min}}^2$. Spatial distributions of particle-scale stress changes for three components \textbf{d}, $p=\frac12(\sigma_{xx}+\sigma_{yy})$, \textbf{e}, $\frac12(\sigma_{xx}-\sigma_{yy})$, \textbf{f}, $\frac12(\sigma_{xy}+\sigma_{yx})$. All of these quantities are measured from $\gamma=0\%$ to $\gamma=0.75\%$, the same as those in main text. Scale bar = 10 d, where d is the diameter of small particle.}
\end{figure}
\newpage
\begin{figure}[h!]
	\centerline{\includegraphics[trim=0cm 0cm 0cm 0cm, width=0.8\linewidth]{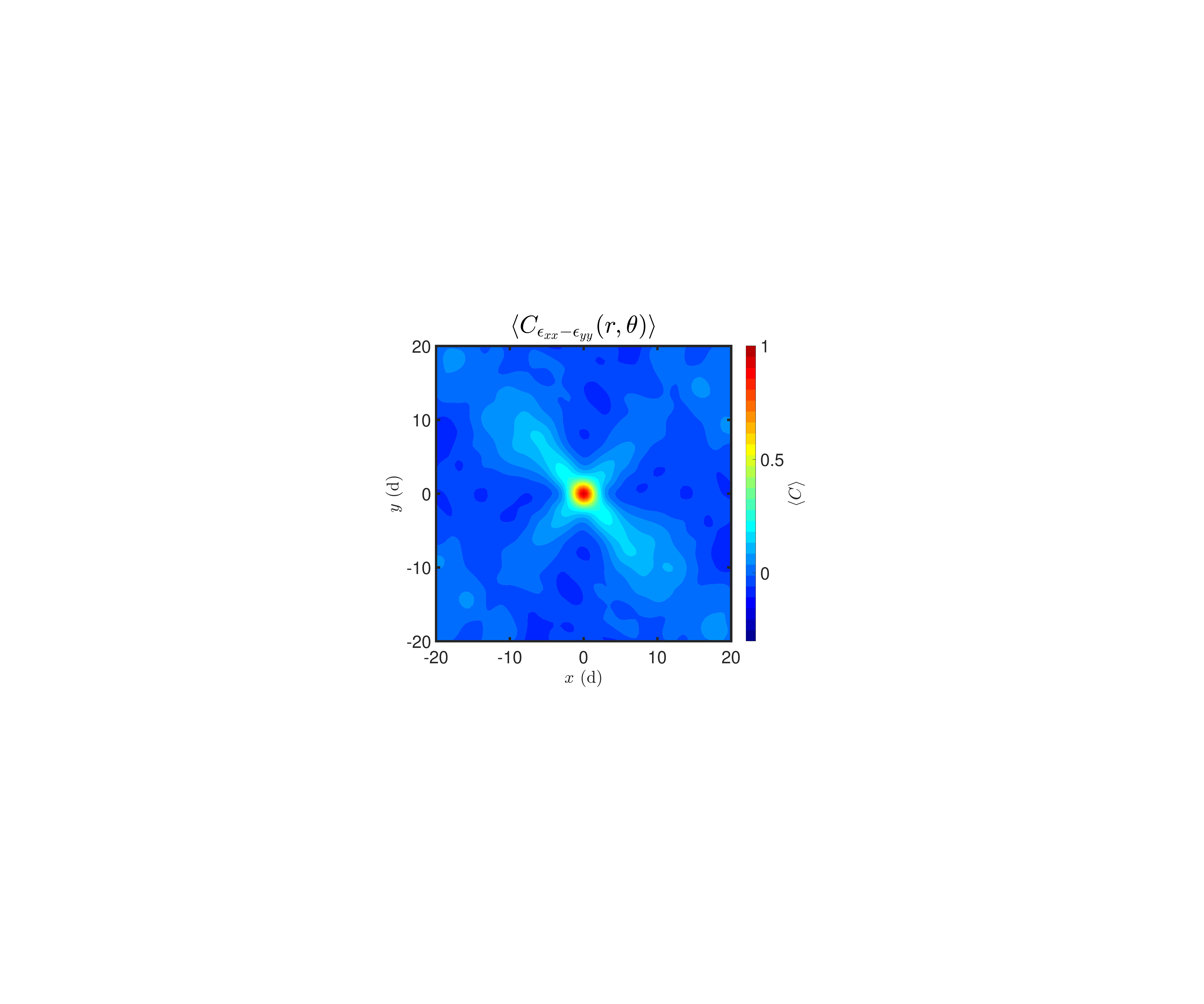}}
	\caption{\textbf{Autocorrelation map of local shear strain $\epsilon_{xx}-\epsilon_{yy}$.} The local shear strain is measured from $\gamma=0\%$ to $\gamma=0.75\%$. The map is averaged over six independent runs. $\langle C \rangle$ denotes the correlation function.}
\end{figure}
\newpage
\begin{figure}[h!]
	\centerline{\includegraphics[trim=0cm 0cm 0cm 0cm, width=0.8\linewidth]{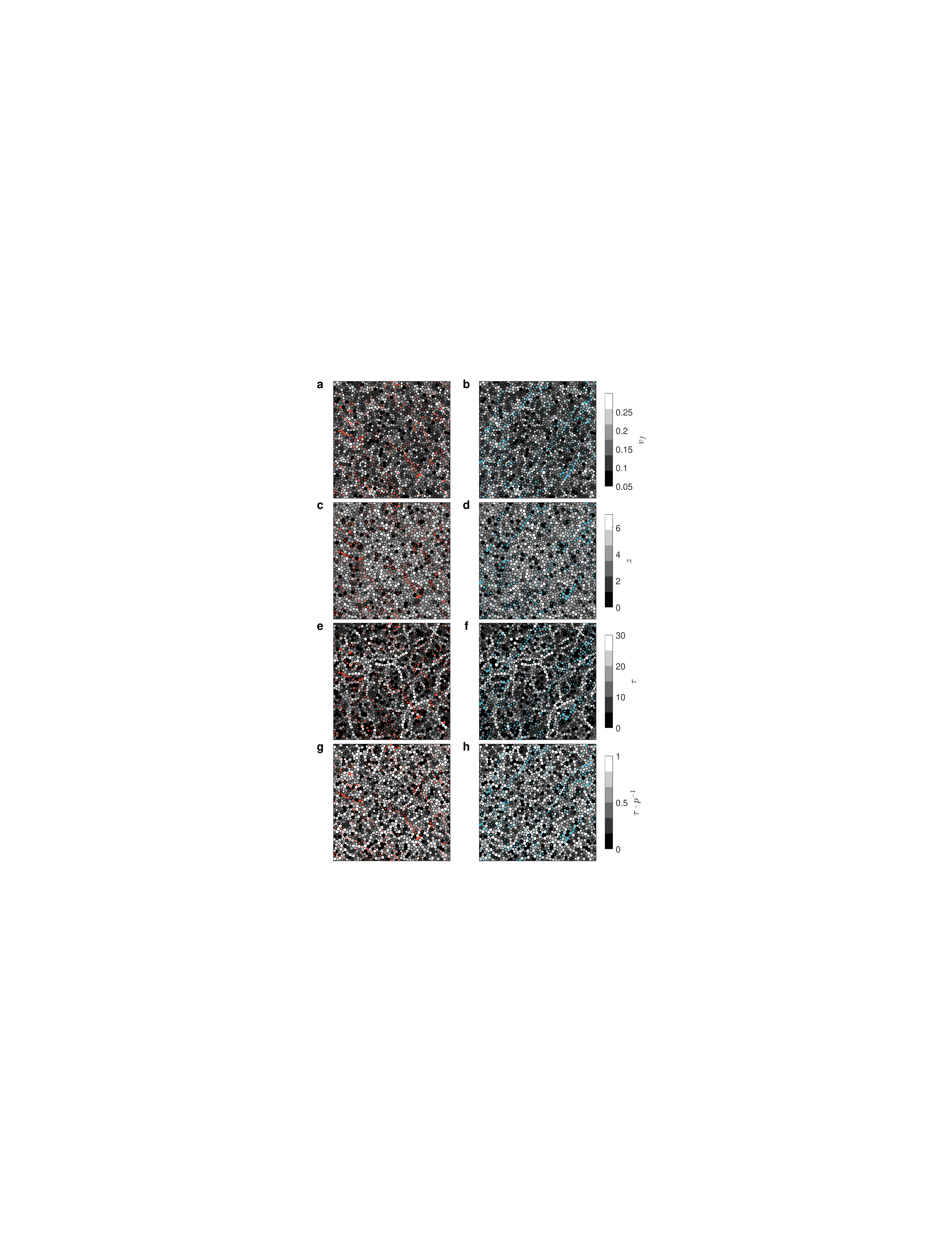}}
	\caption{\textbf{Correlations between particle rotations and other quantities.} \textbf{a}, For counterclockwise rotations $R_+$ and free volume $v_f=(S_i-\pi r_i^2)\cdot(\pi r_i^2)^{-1}$, $S_i$ is the voronoi area of disk $i$ and $r_i$ is its radius. $C(v_f,R_+)=-0.34$. \textbf{b}, For clockwise rotations $R_-$ and $v_f$. $C(v_f,R_-)=-0.19$. \textbf{c}, For $R_+$ and contact number $z$. $C(z,R_+)=0.15$. \textbf{d}, For $R_-$ and $z$. $C(z,R_-)=0.15$. \textbf{e}, For $R_+$ and deviatoric shear stress $\tau$ (one half of the difference between two eigenvalues of a stress tensor). $C(\tau,R_+)=0.08$. \textbf{f}, For $R_-$ and $\tau$. $C(\tau,R_-)=0.16$. \textbf{g}, For $R_+$ and the ratio of the deviatoric shear stress over pressure $\tau\cdot p^{-1}$. $C(\tau\cdot p^{-1},R_+)=-0.15$. \textbf{h}, For $R_-$ and $\tau\cdot p^{-1}$. $C(\tau\cdot p^{-1},R_-)=-0.13$.}
\end{figure}
\newpage
\begin{figure}[h!]
	\centerline{\includegraphics[trim=0cm 0cm 0cm 0cm, width=0.6\linewidth]{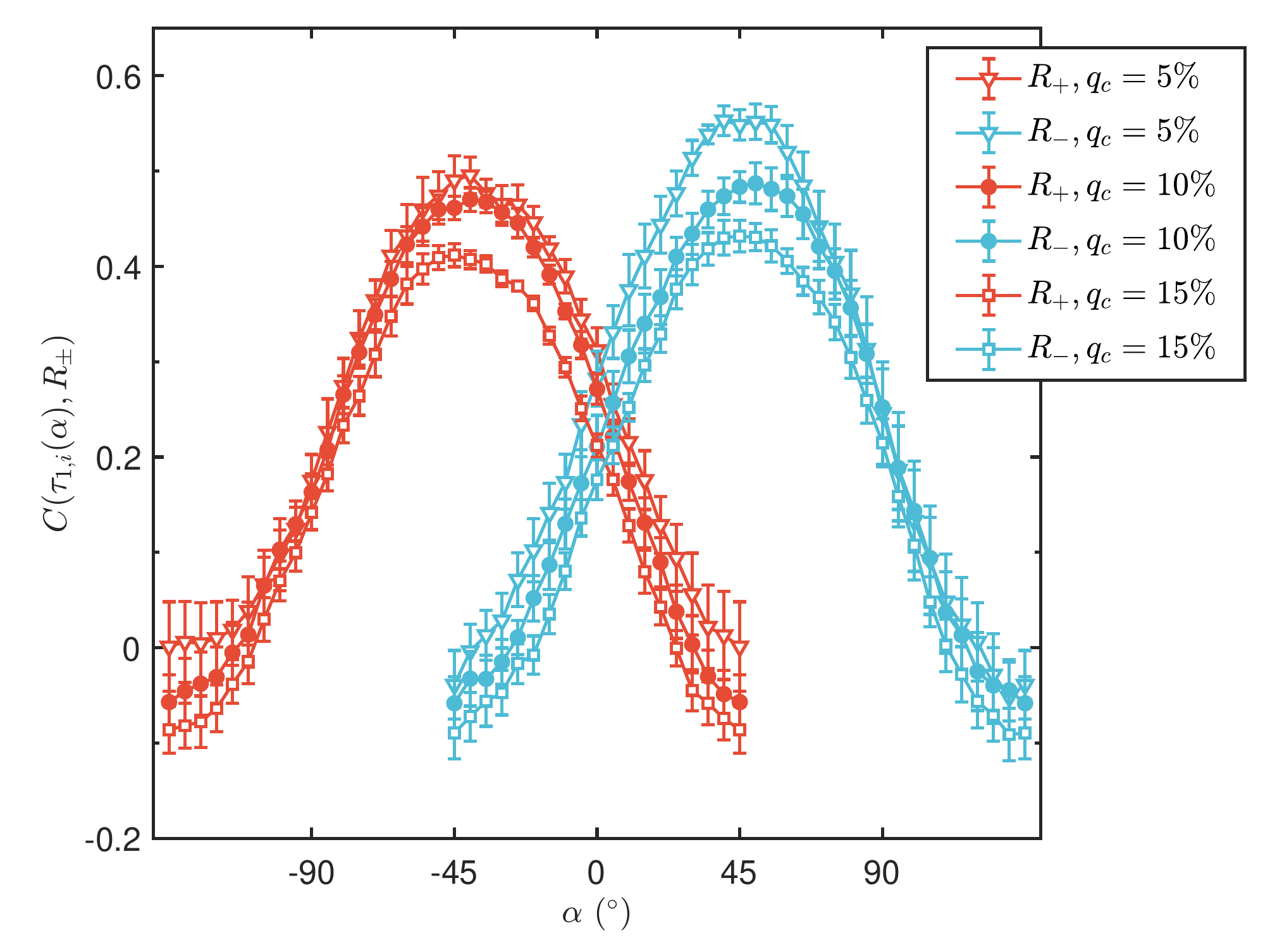}}
	\caption{\textbf{Correlations between particle rotations and $\tau_{1,i}(\alpha)$ versus angle $\alpha$, for different cutoffs $q_c$ of particle rotations.}}
\end{figure}

\begin{figure}[h!]
	\centerline{\includegraphics[trim=0cm 0cm 0cm 0cm, width=0.8\linewidth]{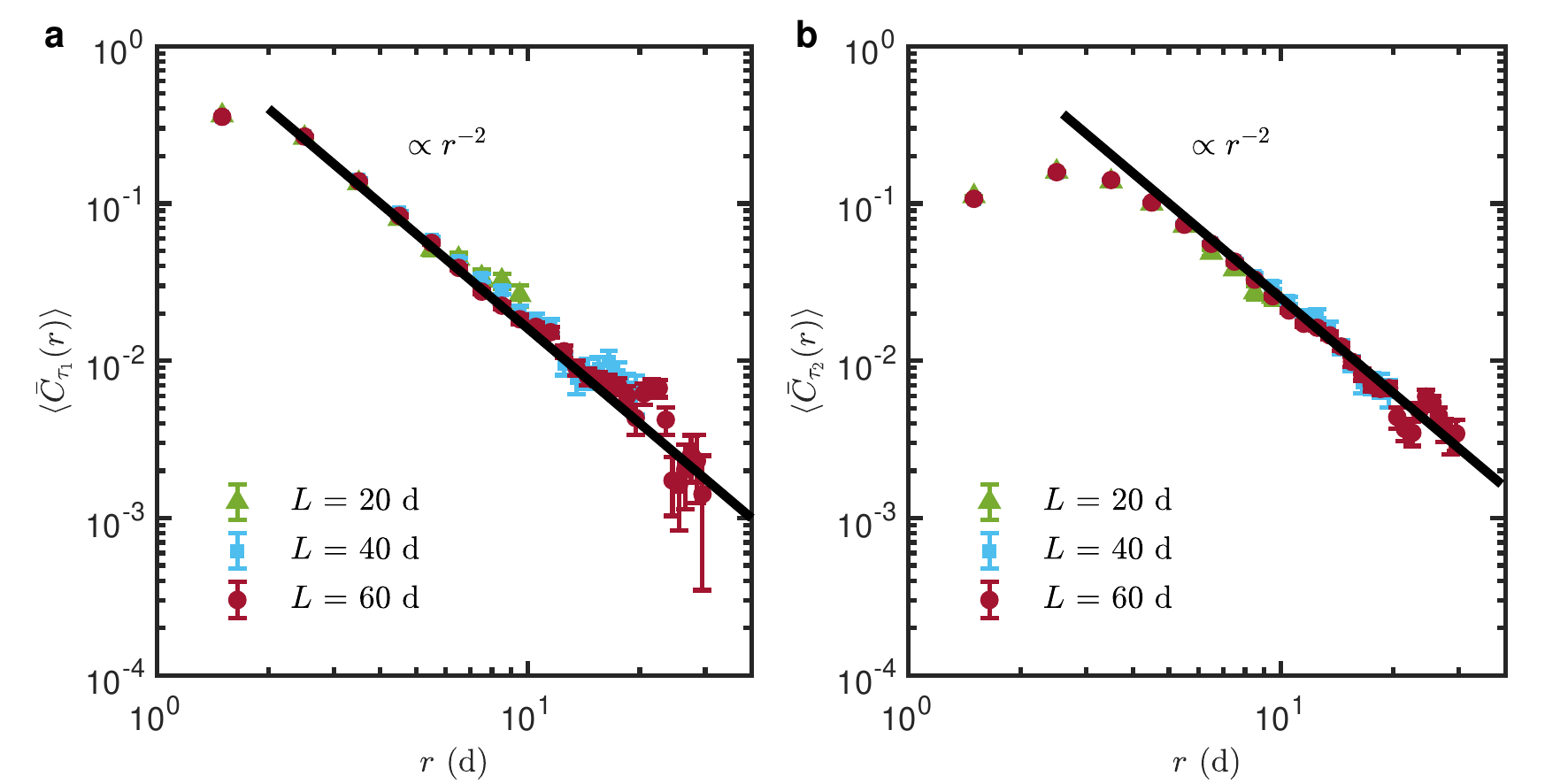}}
	\caption{\textbf{Finite size analyses of stress correlation functions.} Finite size analyses of angle-averaged correlation functions \textbf{a}, $\langle\bar{C}_{\tau_1}(r)\rangle = \pi^{-1}\int_0^{2\pi}\mathrm{d}\theta \cos(2\theta) \langle C_{\tau_1}(r,\theta)\rangle$ and \textbf{b}, $\langle\bar{C}_{\tau_2}(r)\rangle = \pi^{-1}\int_0^{2\pi}\mathrm{d}\theta \cos(4\theta) \langle C_{\tau_2}(r,\theta)\rangle$, for system sizes of $L$ = 20 d, 40 d and 60 d. The black lines indicate a power law of $r^{-2}$, as a guide to the eye.}
\end{figure}


\section*{References}
\begin{thebibliography}{10}
	\expandafter\ifx\csname url\endcsname\relax
	\def\url#1{\texttt{#1}}\fi
	\expandafter\ifx\csname urlprefix\endcsname\relax\def\urlprefix{URL }\fi
	\providecommand{\bibinfo}[2]{#2}
	\providecommand{\eprint}[2][]{\url{#2}}
	
	\bibitem{amon12hotSpots}
	\bibinfo{author}{Amon, A.}, \bibinfo{author}{Nguyen, V.~B.},
	\bibinfo{author}{Bruand, A.}, \bibinfo{author}{Crassous, J.} \&
	\bibinfo{author}{Cl\'{e}ment, E.}
	\newblock \bibinfo{title}{Hot spots in an athermal system}.
	\newblock \emph{\bibinfo{journal}{Phys. Rev. Lett.}}
	\textbf{\bibinfo{volume}{108}}, \bibinfo{pages}{135502}
	(\bibinfo{year}{2012}).
	
	\bibitem{lebouil14granular}
	\bibinfo{author}{Le~Bouil, A.}, \bibinfo{author}{Amon, A.},
	\bibinfo{author}{McNamara, S.} \& \bibinfo{author}{Crassous, J.}
	\newblock \bibinfo{title}{Emergence of cooperativity in plasticity of soft
		glassy materials}.
	\newblock \emph{\bibinfo{journal}{Phys. Rev. Lett.}}
	\textbf{\bibinfo{volume}{112}}, \bibinfo{pages}{246001}
	(\bibinfo{year}{2014}).
	
	\bibitem{denisov16granularAvalanch}
	\bibinfo{author}{Denisov, D.~V.}, \bibinfo{author}{Lorincz, K.~A.},
	\bibinfo{author}{Uhl, J.~T.}, \bibinfo{author}{Dahmen, K.~A.} \&
	\bibinfo{author}{Schall, P.}
	\newblock \bibinfo{title}{Universality of slip avalanches in flowing granular
		matter}.
	\newblock \emph{\bibinfo{journal}{Nat. Commun.}} \textbf{\bibinfo{volume}{7}},
	\bibinfo{pages}{10641} (\bibinfo{year}{2016}).
	
	\bibitem{zhengjie18}
	\bibinfo{author}{Zheng, J.}, \bibinfo{author}{Sun, A.}, \bibinfo{author}{Wang,
		Y.} \& \bibinfo{author}{Zhang, J.}
	\newblock \bibinfo{title}{Energy fluctuations in slowly sheared granular
		materials}.
	\newblock \emph{\bibinfo{journal}{Phys. Rev. Lett.}}
	\textbf{\bibinfo{volume}{121}}, \bibinfo{pages}{248001}
	(\bibinfo{year}{2018}).
	
	\bibitem{nicolas18review}
	\bibinfo{author}{Nicolas, A.}, \bibinfo{author}{Ferrero, E.~E.},
	\bibinfo{author}{Martens, K.} \& \bibinfo{author}{Barrat, J.-L.}
	\newblock \bibinfo{title}{Deformation and flow of amorphous solids: Insights
		from elastoplastic models}.
	\newblock \emph{\bibinfo{journal}{Rev. Mod. Phys.}}
	\textbf{\bibinfo{volume}{90}}, \bibinfo{pages}{045006}
	(\bibinfo{year}{2018}).
	
	\bibitem{albaret16}
	\bibinfo{author}{Albaret, T.}, \bibinfo{author}{Tanguy, A.},
	\bibinfo{author}{Boioli, F.} \& \bibinfo{author}{Rodney, D.}
	\newblock \bibinfo{title}{Mapping between atomistic simulations and eshelby
		inclusions in the shear deformation of an amorphous silicon model}.
	\newblock \emph{\bibinfo{journal}{Phys. Rev. E}} \textbf{\bibinfo{volume}{93}},
	\bibinfo{pages}{053002} (\bibinfo{year}{2016}).
	
	\bibitem{wang12metallicGlass}
	\bibinfo{author}{Wang, W.~H.}
	\newblock \bibinfo{title}{The elastic properties, elastic models and elastic
		perspectives of metallic glasses}.
	\newblock \emph{\bibinfo{journal}{Prog. Mater. Sci.}}
	\textbf{\bibinfo{volume}{57}}, \bibinfo{pages}{487--656}
	(\bibinfo{year}{2012}).
	
	\bibitem{chikkadi11strainCorrelation}
	\bibinfo{author}{Chikkadi, V.}, \bibinfo{author}{Wegdam, G.},
	\bibinfo{author}{Bonn, D.}, \bibinfo{author}{Nienhuis, B.} \&
	\bibinfo{author}{Schall, P.}
	\newblock \bibinfo{title}{Long-range strain correlations in sheared colloidal
		glasses}.
	\newblock \emph{\bibinfo{journal}{Phys. Rev. Lett.}}
	\textbf{\bibinfo{volume}{107}}, \bibinfo{pages}{198303}
	(\bibinfo{year}{2011}).
	
	\bibitem{jensen14quadrupole}
	\bibinfo{author}{Jensen, K.~E.}, \bibinfo{author}{Weitz, D.~A.} \&
	\bibinfo{author}{Spaepen, F.}
	\newblock \bibinfo{title}{Local shear transformations in deformed and quiescent
		hard-sphere colloidal glasses}.
	\newblock \emph{\bibinfo{journal}{Phys. Rev. E}} \textbf{\bibinfo{volume}{90}},
	\bibinfo{pages}{042305--042305} (\bibinfo{year}{2014}).
	
	\bibitem{illing16supercooledLiquids}
	\bibinfo{author}{Illing, B.} \emph{et~al.}
	\newblock \bibinfo{title}{Strain pattern in supercooled liquids}.
	\newblock \emph{\bibinfo{journal}{Phys. Rev. Lett.}}
	\textbf{\bibinfo{volume}{117}}, \bibinfo{pages}{208002}
	(\bibinfo{year}{2016}).
	
	\bibitem{desmond15emulsions}
	\bibinfo{author}{Desmond, K.~W.} \& \bibinfo{author}{Weeks, E.~R.}
	\newblock \bibinfo{title}{Measurement of stress redistribution in flowing
		emulsions}.
	\newblock \emph{\bibinfo{journal}{Phys. Rev. Lett.}}
	\textbf{\bibinfo{volume}{115}}, \bibinfo{pages}{098302}
	(\bibinfo{year}{2015}).
	
	\bibitem{kabla09Foam}
	\bibinfo{author}{Kabla, A.} \& \bibinfo{author}{Debr\'{e}geas, G.}
	\newblock \bibinfo{title}{Local stress relaxation and shear banding in a dry
		foam under shear}.
	\newblock \emph{\bibinfo{journal}{Phys. Rev. Lett.}}
	\textbf{\bibinfo{volume}{90}}, \bibinfo{pages}{258303}
	(\bibinfo{year}{2003}).
	
	\bibitem{terzaghi96soilMechanics}
	\bibinfo{author}{Terzaghi, K.}, \bibinfo{author}{Peck, R.~B.} \&
	\bibinfo{author}{Mesri, G.}
	\newblock \emph{\bibinfo{title}{Soil mechanics in engineering practice}}
	(\bibinfo{publisher}{John Wiley \& Sons}, \bibinfo{year}{1996}).
	
	\bibitem{scholz19earthquakeBook}
	\bibinfo{author}{Scholz, C.~H.}
	\newblock \emph{\bibinfo{title}{The mechanics of earthquakes and faulting}}
	(\bibinfo{publisher}{Cambridge University Press}, \bibinfo{year}{2019}).
	
	\bibitem{langer01STZ}
	\bibinfo{author}{Langer, J.}
	\newblock \bibinfo{title}{Microstructural shear localization in plastic
		deformation of amorphous solids}.
	\newblock \emph{\bibinfo{journal}{Phys. Rev. E}} \textbf{\bibinfo{volume}{64}},
	\bibinfo{pages}{011504} (\bibinfo{year}{2001}).
	
	\bibitem{maloney06}
	\bibinfo{author}{Maloney, C.~E.} \& \bibinfo{author}{Lema\^{i}tre, A.}
	\newblock \bibinfo{title}{Amorphous systems in athermal, quasistatic shear}.
	\newblock \emph{\bibinfo{journal}{Phys. Rev. E}} \textbf{\bibinfo{volume}{74}},
	\bibinfo{pages}{016118} (\bibinfo{year}{2006}).
	
	\bibitem{tanguy06}
	\bibinfo{author}{Tanguy, A.}, \bibinfo{author}{Leonforte, F.} \&
	\bibinfo{author}{Barrat, J.~L.}
	\newblock \bibinfo{title}{Plastic response of a 2d lennard-jones amorphous
		solid: Detailed analysis of the local rearrangements at very slow strain
		rate}.
	\newblock \emph{\bibinfo{journal}{Eur. Phys. J. E}}
	\textbf{\bibinfo{volume}{20}}, \bibinfo{pages}{355--364}
	(\bibinfo{year}{2006}).
	
	\bibitem{dasgupta12energyMinimization}
	\bibinfo{author}{Dasgupta, R.}, \bibinfo{author}{Hentschel, H. G.~E.} \&
	\bibinfo{author}{Procaccia, I.}
	\newblock \bibinfo{title}{Microscopic mechanism of shear bands in amorphous
		solids}.
	\newblock \emph{\bibinfo{journal}{Phys. Rev. Lett.}}
	\textbf{\bibinfo{volume}{109}}, \bibinfo{pages}{255502}
	(\bibinfo{year}{2012}).
	
	\bibitem{eshelby57}
	\bibinfo{author}{Eshelby, J.~D.}
	\newblock \bibinfo{title}{The determination of the elastic field of an
		ellipsoidal inclusion, and related problems}.
	\newblock \emph{\bibinfo{journal}{Proc. R. Soc. Lond.}}
	\textbf{\bibinfo{volume}{241}}, \bibinfo{pages}{376--396}
	(\bibinfo{year}{1957}).
	
	\bibitem{chowdhury16harrowell}
	\bibinfo{author}{Chowdhury, S.}, \bibinfo{author}{Abraham, S.},
	\bibinfo{author}{Hudson, T.} \& \bibinfo{author}{Harrowell, P.}
	\newblock \bibinfo{title}{Long range stress correlations in the inherent
		structures of liquids at rest}.
	\newblock \emph{\bibinfo{journal}{J. Chem. Phys.}}
	\textbf{\bibinfo{volume}{144}}, \bibinfo{pages}{124508}
	(\bibinfo{year}{2016}).
	
	\bibitem{lemaitre17inherentStress}
	\bibinfo{author}{Lema\^{i}tre, A.}
	\newblock \bibinfo{title}{Inherent stress correlations in a quiescent
		two-dimensional liquid: Static analysis including finite-size effects}.
	\newblock \emph{\bibinfo{journal}{Phys. Rev. E}} \textbf{\bibinfo{volume}{96}},
	\bibinfo{pages}{052101} (\bibinfo{year}{2017}).
	
	\bibitem{maier17}
	\bibinfo{author}{Maier, M.}, \bibinfo{author}{Zippelius, A.} \&
	\bibinfo{author}{Fuchs, M.}
	\newblock \bibinfo{title}{Emergence of long-ranged stress correlations at the
		liquid to glass transition}.
	\newblock \emph{\bibinfo{journal}{Phys. Rev. Lett.}}
	\textbf{\bibinfo{volume}{119}}, \bibinfo{pages}{265701}
	(\bibinfo{year}{2017}).
	
	\bibitem{lemaitre18theory}
	\bibinfo{author}{Lema\^{i}tre, A.}
	\newblock \bibinfo{title}{Stress correlations in glasses}.
	\newblock \emph{\bibinfo{journal}{J. Chem. Phys.}}
	\textbf{\bibinfo{volume}{149}}, \bibinfo{pages}{104107}
	(\bibinfo{year}{2018}).
	
	\bibitem{henkes2009fieldTheory}
	\bibinfo{author}{Henkes, S.} \& \bibinfo{author}{Chakraborty, B.}
	\newblock \bibinfo{title}{Statistical mechanics framework for static granular
		matter}.
	\newblock \emph{\bibinfo{journal}{Phys. Rev. E}} \textbf{\bibinfo{volume}{79}},
	\bibinfo{pages}{061301} (\bibinfo{year}{2009}).
	
	\bibitem{degiuli18PRE}
	\bibinfo{author}{DeGiuli, E.}
	\newblock \bibinfo{title}{Edwards field theory for glasses and granular
		matter}.
	\newblock \emph{\bibinfo{journal}{Phys. Rev. E}} \textbf{\bibinfo{volume}{98}},
	\bibinfo{pages}{033001} (\bibinfo{year}{2018}).
	
	\bibitem{degiuli18PRL}
	\bibinfo{author}{DeGiuli, E.}
	\newblock \bibinfo{title}{Field theory for amorphous solids}.
	\newblock \emph{\bibinfo{journal}{Phys. Rev. Lett.}}
	\textbf{\bibinfo{volume}{121}}, \bibinfo{pages}{118001}
	(\bibinfo{year}{2018}).
	
	\bibitem{majmudar05nature}
	\bibinfo{author}{Majmudar, T.~S.} \& \bibinfo{author}{Behringer, R.~P.}
	\newblock \bibinfo{title}{Contact force measurements and stress-induced
		anisotropy in granular materials}.
	\newblock \emph{\bibinfo{journal}{Nature}} \textbf{\bibinfo{volume}{435}},
	\bibinfo{pages}{1079--82} (\bibinfo{year}{2005}).
	
	\bibitem{ostojic06forceNetwork}
	\bibinfo{author}{Ostojic, S.}, \bibinfo{author}{Somfai, E.} \&
	\bibinfo{author}{Nienhuis, B.}
	\newblock \bibinfo{title}{Scale invariance and universality of force networks
		in static granular matter}.
	\newblock \emph{\bibinfo{journal}{Nature}} \textbf{\bibinfo{volume}{439}},
	\bibinfo{pages}{828--830} (\bibinfo{year}{2006}).
	
	\bibitem{falk98langer}
	\bibinfo{author}{Falk, M.} \& \bibinfo{author}{Langer, J.}
	\newblock \bibinfo{title}{Dynamics of viscoplastic deformation in amorphous
		solids}.
	\newblock \emph{\bibinfo{journal}{Phys. Rev. E}} \textbf{\bibinfo{volume}{57}},
	\bibinfo{pages}{7192} (\bibinfo{year}{1998}).
	
	\bibitem{patinet16localYieldStress}
	\bibinfo{author}{Patinet, S.}, \bibinfo{author}{Vandembroucq, D.} \&
	\bibinfo{author}{Falk, M.~L.}
	\newblock \bibinfo{title}{Connecting local yield stresses with plastic activity
		in amorphous solids}.
	\newblock \emph{\bibinfo{journal}{Phys. Rev. Lett.}}
	\textbf{\bibinfo{volume}{117}}, \bibinfo{pages}{045501}
	(\bibinfo{year}{2016}).
	
	\bibitem{goldhirsch02CG}
	\bibinfo{author}{Goldhirsch, I.} \& \bibinfo{author}{Goldenberg, C.}
	\newblock \bibinfo{title}{On the microscopic foundations of elasticity}.
	\newblock \emph{\bibinfo{journal}{Eur. Phys. J. E}}
	\textbf{\bibinfo{volume}{9}}, \bibinfo{pages}{245--251}
	(\bibinfo{year}{2002}).
	
	\bibitem{lois2009}
	\bibinfo{author}{Lois, G.} \emph{et~al.}
	\newblock \bibinfo{title}{Stress correlations in granular materials: An
		entropic formulation}.
	\newblock \emph{\bibinfo{journal}{Phys. Rev. E}} \textbf{\bibinfo{volume}{80}},
	\bibinfo{pages}{060303} (\bibinfo{year}{2009}).
	
	\bibitem{schirmacher13bosonPeak}
	\bibinfo{author}{Schirmacher, W.}
	\newblock \bibinfo{title}{The boson peak}.
	\newblock \emph{\bibinfo{journal}{Phys. Status Solidi B}}
	\textbf{\bibinfo{volume}{250}}, \bibinfo{pages}{937--943}
	(\bibinfo{year}{2013}).
	
	\bibitem{wang18bosonPeak}
	\bibinfo{author}{Wang, Y.}, \bibinfo{author}{Hong, L.}, \bibinfo{author}{Wang,
		Y.}, \bibinfo{author}{Schirmacher, W.} \& \bibinfo{author}{Zhang, J.}
	\newblock \bibinfo{title}{Disentangling boson peaks and van hove singularities
		in a model glass}.
	\newblock \emph{\bibinfo{journal}{Phys. Rev. B}} \textbf{\bibinfo{volume}{98}},
	\bibinfo{pages}{174207} (\bibinfo{year}{2018}).
	
	\bibitem{ohern03Jamming}
	\bibinfo{author}{O'Hern, C.~S.}, \bibinfo{author}{Silbert, L.~E.},
	\bibinfo{author}{Liu, A.~J.} \& \bibinfo{author}{Nagel, S.~R.}
	\newblock \bibinfo{title}{Jamming at zero temperature and zero applied stress:
		The epitome of disorder}.
	\newblock \emph{\bibinfo{journal}{Phys. Rev. E}} \textbf{\bibinfo{volume}{68}},
	\bibinfo{pages}{011306} (\bibinfo{year}{2003}).
	
	\bibitem{daniels17photoelasticity}
	\bibinfo{author}{Daniels, K.~E.}, \bibinfo{author}{Kollmer, J.~E.} \&
	\bibinfo{author}{Puckett, J.~G.}
	\newblock \bibinfo{title}{Photoelastic force measurements in granular
		materials}.
	\newblock \emph{\bibinfo{journal}{Rev. Sci. Instrum.}}
	\textbf{\bibinfo{volume}{88}}, \bibinfo{pages}{051808}
	(\bibinfo{year}{2017}).
	
	\bibitem{zadeh19}
	\bibinfo{author}{Abed~Zadeh, A.} \emph{et~al.}
	\newblock \bibinfo{title}{Enlightening force chains: a review of
		photoelasticimetry in granular matter}.
	\newblock \emph{\bibinfo{journal}{Granul. Matter}}
	\textbf{\bibinfo{volume}{21}}, \bibinfo{pages}{83} (\bibinfo{year}{2019}).
	
\end{thebibliography}
\end{document}